\documentstyle[aps,prl,twocolumn,floats,epsf]{revtex}

\begin{document}
\draft
\twocolumn[\hsize\textwidth\columnwidth\hsize\csname@twocolumnfalse\endcsname
\title{
Crossover scaling from classical to nonclassical critical behavior.
}
\author{Andrea Pelissetto, Paolo Rossi, and Ettore Vicari}
\address{Dipartimento di Fisica dell'Universit\`a 
and I.N.F.N.,I-56126 Pisa, Italy}

\maketitle

\begin{abstract}
We study the crossover between classical and nonclassical critical behaviors.
The critical crossover limit is  driven by 
the Ginzburg number $G$.  The corresponding scaling functions
are universal with respect to any
possible microscopic mechanism  which can vary $G$, such
as changing the range or the strength of the interactions.
The critical crossover describes the unique flow from the
unstable Gaussian to the stable nonclassical fixed point. 
The scaling functions are related to the 
continuum renormalization-group functions.
We show these features explicitly in the large-$N$ limit of 
the O$(N)$ $\phi^4$ model. We also show that the effective susceptibility 
exponent is nonmonotonic in the low-temperature phase of the 
three-dimensional Ising model.

\end{abstract}


]

\narrowtext

Motivated by various experimental results
(see e.g. Refs.~\cite{C-D,Fisher-prl,A-P-K-S,F-L}), 
there has been recently 
a revived interest in understanding crossover phenomena
driven by the effective range of the interactions.
If the interactions have a finite range $R$, in the limit in which 
the reduced temperature $t$ goes to zero,
the system shows the standard short-range nonclassical behavior.
According to the Ginzburg criterion \cite{Ginzburg}  this occurs when 
$t\ll G$ where $t$ and $G$ are respectively the reduced temperature
and the Ginzburg number.
On the other hand, in the opposite limit $t\gg G$  the system shows a
classical Gaussian behavior. In the intermediate
region one observes 
a crossover between these two behaviors.
{}From the point of view
of the Wilson renormalization-group (RG) theory, 
this crossover phenomenon is generally explained by the competition of 
two fixed points: the Gaussian fixed point
and the nonclassical fixed point  that determines the asymptotic 
behavior in the neighborhood of criticality. 

These crossover phenomena are of great importance for the understanding of
critical phenomena occurring in physical systems 
(see e.g. Refs.~\cite{Fisher-prl,D-B,A-P-K-S}).
Fisher~\cite{Fisher-prl} discussed experiments
on micellar solutions~\cite{C-D} and argued that the apparently 
nonuniversal results of the critical exponents
may be explained in terms of a crossover behavior
driven by the effective range of the interactions.
Crossover phenomena are also observed in 
experimental data for the susceptibility of fluids
and liquid mixtures~\cite{A-P-K-S}, and in polymer melts \cite{D-B}. 
Some understanding of the 
crossover problem is provided by field-theoretic 
calculations (see e.g. 
Refs.\cite{S-F,N-B,B-B-l,N-A,B-B,B-B-M-N,C-A-S,A-K-S-T,B-K}).

The most important issue
concerning crossover phenomena is whether one can define 
scaling functions that are universal and that describe the crossover 
between the classical (Gaussian) and the nonclassical behaviour.
In order to give an answer to this question
one must clarify the kind of crossover one is considering. 
Varying $t$, 
while keeping the Ginzburg number $G$ fixed, 
one observes  a crossover  between the critical 
and the noncritical behavior.
This is obviously non-universal  
and for $t\lesssim G$ it is described by the
nonuniversal Wegner expansion~\cite{notawe}. 
As pointed out by Bagnuls and Bervillier~\cite{B-B-prl}
(see also \cite{A-P-K-S-2}) a universal behaviour can only be obtained 
in a properly defined critical limit. 
In other words, a universal behaviour can be observed 
only if $t\ll 1$, i.e. the correlation
length is large, in the whole crossover region.
The problem was  properly formulated by
Luijten, Bl\"ote and Binder \cite{L-B-B-pre,L-B-B-prl}, 
who argued that a  universal
crossover description  can be obtained if one considers the simultaneous 
limits $t\to 0$, $R\to \infty$ keeping the product between $t$ and
an appropriate power of $R$ fixed.
A Wilson RG analysis \cite{L-B-B-pre} indicates 
that this limit is non trivial and interpolates 
between the mean-field and the standard short-range behaviour. 
These ideas have  been confirmed 
numerically in the two-dimensional Ising model \cite{L-B-B-prl}.
We will refer to the above limit as critical crossover.
Thus the critical crossover is a crossover from the 
critical nonclassical behavior to the
critical  classical behavior.
Extending the arguments
of Ref.~\cite{L-B-B-prl}, we define the appropriate crossover limit
in the whole $(t,h)$ plane introducing a
magnetic Ginzburg number $G_h$, such that the system
shows classical behaviour for $h \gg G_h$ and the standard short-range
behaviour in the opposite case. 

The main point is that in the critical crossover  region
the key role is played by the Ginzburg number $G$:
the range of the interaction is only one of the possible microscopic
parameters controlling  $G$. Any other mechanism leading
to a change of $G$ can  give rise to the same critical crossover
when the appropriate limit is considered. 
We show that the critical crossover 
driven by $R$ can be reproduced starting from
a standard $\phi^4$ theory with short-range interactions
and taking an appropriate limit
of the theory when the bare four-point coupling goes to zero.
The critical crossover functions expressed in terms of the renormalized
coupling are related to the standard continuum RG functions. 

The several open questions on the crossover behavior 
call for a theoretical laboratory where the 
various conjectures can be verified analytically.
The O($N$) vector model in the large-$N$ limit is ideal for this purpose.
Indeed, although it maintains many nontrivial features of the
theory, it allows us to perform exact calculations,
and therefore an exact verification of the conjectures
on the crossover phenomena. 
Starting from a lattice O($N$) model with long-range interactions,
we calculate the crossover functions for $2<d<4$,
and discuss their universality in
the critical crossover region. We derive the 
equation of state that provides a complete description of the 
crossover region. 

For the sake of definiteness,
we consider the 
$d$-dimensional lattice model defined by the Hamiltonian
\begin{eqnarray}
{\cal H}=&&\sum_{i,j} \case{1}{2}J(\vec{x}_i-\vec{x}_j) 
\left[\phi(\vec{x}_i) - \phi(\vec{x}_j)\right]^2 
\nonumber \\
&&+\sum_i \left[ \case{1}{2}  r\phi(\vec{x}_i)^2 + 
\case{1}{4!}u \phi(\vec{x}_i)^4 - h \phi(\vec{x}_i)\right] ,
\label{lham}
\end{eqnarray}
where $\phi(\vec{x}_i)$ are $N$-dimensional vectors.
The spin-spin interaction $J(\vec{x})$ has finite range $R$
defined by
\begin{equation}
R^2 =\;
 {1\over 2d} {\sum_{\vec{x}} x^2 J(\vec{x}) \over \sum_{\vec{x}} J(\vec{x})}.
\label{defR}
\end{equation}
The specific form of $J(\vec{x})$ is irrelevant for our discussion.
The normalization of $J(\vec{x})$ is chosen so that 
its Fourier transform $\Pi \equiv \Pi(k,R)$
has the low-momentum behavior $k^2 + O(k^4)$. 
The Ginzburg criterion~\cite{Ginzburg} applied to the model 
(\ref{lham}) tells us that the theory has
a nonclassical critical behavior when
\begin{equation}
t\equiv r-r_c \ll G\propto u^{2/(4-d)}.
\label{Gu}
\end{equation}

In order to study the long-range limit, it is 
convenient to perform  a field rescaling with a
corresponding rescaling of the 
Hamiltonian parameters:
\begin{equation}
\overline{\phi} \equiv R^{-1}\phi, \quad
\overline{t}\equiv R^2 t,\quad \overline{u}\equiv  R^4 u,\quad
\overline{h}\equiv  R h.
\label{bres}
\end{equation}
Keeping $R$ finite, the critical behavior for 
$\overline{t}\to 0$ and $\overline{h}\to 0$ is nonclassical. 
For example in the large-$N$ 
limit the critical exponents are $\gamma=2\nu=2/(d-2)$, $\eta=0$,
$\beta=1/2$, $\delta=(d+2)/(d-2)$. 
When $R\rightarrow\infty$ one expects 
a mean-field critical behavior, characterized by the exponents
$\gamma=2\nu=1$, $\eta=0$, $\beta=1/2$, $\delta=3$. 
This change implies a singular dependence on $R$ of the critical amplitudes,
which has been derived in Refs.~\cite{M-B,L-B-B-pre}.
For example the asymptotic behaviors 
of the magnetization, of the magnetic susceptibility and of the 
correlation length, 
for $\overline{t}\to 0$ 
(with $h=0$), 
are expected to be~\cite{M-B,L-B-B-pre}
\begin{eqnarray}
&&\overline{\chi}\equiv \sum_i 
\langle \overline{\phi}_0\cdot \overline{\phi}_i\rangle
 \propto \overline{t}^{-\gamma} R^{2d(1-\gamma)/(4-d)},\label{chisc}\\
&&\overline{\xi}^2 \equiv {1\over 2d\overline{\chi}R^2} \sum_i 
x^2_i\langle \overline{\phi}_0\cdot \overline{\phi}_i\rangle
 \propto \overline{t}^{-2\nu} R^{2d(1-2\nu)/(4-d)},\label{xisc}\\
&&\overline{M}\equiv \langle \overline{\phi}\rangle
\propto \overline{t}^\beta R^{d(2\beta-1)/(4-d)}.\label{msc}
\label{rscaling}
\end{eqnarray}
Moreover, using the Wilson re\-nor\-ma\-liza\-tion-group 
approach of Ref.~\cite{L-B-B-pre}, we find that at $t=0$ 
\begin{equation}
\overline{M} \propto \overline{h}^{1/\delta} R^{d(3/\delta-1)/(4-d)}.
\label{rhscaling}
\end{equation}
In order to describe the critical crossover from the nonclassical
to the classical behavior as driven by the range of the interaction,
i.e.  keeping $\overline{u}$ fixed,
it is convenient to introduce a new Ginzburg number
associated with $\overline{t}$~\cite{L-B-B-pre}: 
\begin{equation}
\overline{G}\equiv R^{-2d/(4-d)}\left( N\overline{u}\right)^{2/(4-d)}
\propto R^2 G
\label{GB}
\end{equation}
(we have introduced the $N$-dependence only to make the
large-$N$ limit more transparent).
Thus the comparison of  $\overline{t}$ with 
$\overline{G}\propto R^{-2d/(4-d)}$ tells us whether
the critical behavior is nonclassical ($\overline{t}\ll \overline{G}$)
or classical ($\overline{t}\gg \overline{G}$). 

Following Refs.~\cite{L-B-B-pre,L-B-B-prl}, 
one may introduce the rescaled reduced temperature 
\begin{equation}
\widetilde{t} \equiv \overline{t}/\overline{G}
\propto t/G,
\label{tildet}
\end{equation}
and consider the limit $\overline{t}\rightarrow 0$ and
$R\rightarrow \infty$ keeping $\widetilde{t}$ fixed.
When $\widetilde{t}\rightarrow 0$ ($\widetilde{t}\rightarrow \infty$) 
the nonclassical (classical) critical behavior should be recovered.
Extending the RG analysis of Ref.~\cite{L-B-B-pre}
to the line $t = 0$, 
we also introduce a magnetic Ginzburg number
$G_h\propto u^{(d+2)/[2(4-d)]}$ and 
\begin{equation}
\overline{G}_h\equiv R^{-3d/(4-d)}\left( 
N\overline{u}\right)^{(d+2)/[2(4-d)]} \propto R G_h.
\label{gh}
\end{equation}
$\overline{G}_h$ tells us, in the presence of a magnetic field, 
in which regime we are:
nonclassical when $\overline{h}\ll \overline{G}_h$ and 
classical when $\overline{h}\gg \overline{G}_h$.
Correspondingly we define a rescaled magnetic field 
\begin{equation}
\widetilde{h}\equiv \overline{h}\,/\,\overline{G}_h \propto h\,/G_h,
\label{tildeh}
\end{equation}
and study the behavior of the theory when $\overline{h}\rightarrow 0$, 
$R\to\infty$ with
$\widetilde{h}$ fixed. The nonclassical (classical) behavior is 
obtained in the limit 
$\widetilde{h}\rightarrow 0$ ($\widetilde{h}\rightarrow\infty$).

The relations (\ref{chisc}) and (\ref{xisc}) suggest
the following scaling behaviors in the critical crossover  limit
for $h=0$
\begin{eqnarray}
&\widetilde{\chi}\equiv \overline{\chi} \,\overline{G} \approx  
F_\chi(\widetilde{t}\,)\propto \chi\ G,\label{fchi}\\
&\widetilde{\xi}^2\equiv \overline{\xi}^2 \,\overline{G} \approx  
F_{\xi^2}(\widetilde{t}\,)\propto \xi^2\ G.\label{fxi}
\label{scalfunc}
\end{eqnarray}
From Eq.~(\ref{rhscaling}) one obtains 
\begin{equation}
\widetilde{M}\equiv \overline{M} \,\overline{G}\,/\,\overline{G}_h
\approx  F_M(\widetilde{h}\,)\propto M G/G_h \quad {\rm for} 
\quad t=0.
\label{fm}
\end{equation}
$F_\chi (x)$, $F_{\xi^2}(x)$ and $F_M(x)$ are expected to behave as 
$F_\chi(x)\sim x^{-\gamma}$, $F_{\xi^2} (x)\sim x^{-2\nu}$ and $F_M(x)\sim x^{1/\delta}$
for $x\to 0$, and
$F_\chi(x)\sim x^{-1}$, $F_{\xi^2} (x)\sim x^{-1}$ and $F_M(x)\sim x^{1/3}$
for $x\to\infty$.
The corrections to these 
asymptotic behaviours should be  controlled by 
the corresponding leading correction-to-scaling exponents $\Delta$ 
(see e.g. Ref.~\cite{Fisher-prl} and references therein).

It is crucial to notice that in the crossover region the relevant
new scale is provided by the Ginzburg number $G$,
and the critical crossover limit can be expressed
in terms of $G$ (or $G_h$) only, i.e. without the explicit use of $R$.
The range of the interactions represents
a physical way to vary $G$ according to Eq.~(\ref{GB}),
but it is not the only way. The critical crossover scaling
functions are therefore expected to be universal
with respect to the microscopic ways one uses to control and vary
$G$. In other words, the critical crossover describes
the unique flow from the unstable
Gaussian to the stable nonclassical fixed point.
Starting from a $\phi^4$ short-range theory, for instance from the
Hamiltonian (\ref{lham}) with $R=1$ fixed,
one may use the bare four-point coupling $u$ to
vary $G$ according to Eq.~(\ref{Gu}).
One then recovers the critical crossover behavior
in the limit $u\to 0$, $t\to 0$ with
$\widetilde{t}\equiv t/G \propto t u^{-2/(4-d)}$ fixed.
In the context of the statistical approach to polymers the
limit $u\to 0$ keeping $\widetilde{t}\propto tu^{-2/(4-d)}$ fixed
is essentially equivalent to the so-called
two-parameter model (see e.g. Refs.~\cite{M-N,Sokal,B-N}
and references therein). In this limit
the crossover functions can be computed 
in the standard continuum $\phi^4$ theory 
\cite{B-B-l,B-B,B-B-M-N}.
A dimensional analysis shows that 
(using the subtracted bare mass and removing the cutoff)
finite results can be  obtained in terms of the dimensionless
variable $u/t^{2-d/2} =\widetilde{t}^{d/2-2}$, and no further
limiting procedure is required. It is useful to introduce a 
renormalized coupling $g$.
Changing variables from $\widetilde{t}$ to $g$, one may then show
that the critical crossover functions expressed
in terms of $g$ are related to
the standard continuum RG functions. They describe
the physics of strongly correlated systems
in the whole range between the classical and nonclassical
critical point, in terms of a single physical parameter
measuring the ratio between the interaction scale and the correlation
scale.
The critical crossover functions for physically interesting systems
are well studied \cite{B-B-l,B-B,B-B-M-N}
in the fixed-dimension expansion when $d=3$. 

To check explicitly these ideas, let us
consider the large-$N$ limit. Before any rescaling 
the following saddle-point equations hold
\begin{eqnarray}
&& M^2 + {6\over u}( t - m^2) =
N\int {d^dk\over (2\pi)^d} {m^2\over \Pi (\Pi + m^2) } ,
\label{sapo} \\
&& M = {h\over m^2},
\label{M-h-beta-relation}
\end{eqnarray}
where
$M\equiv \langle\phi\rangle$ is the magnetization and
$\xi= 1/m$ a dynamically determined
length scale.
For $h=0$ 
\begin{equation}
\overline{\chi} = {N/\overline{m}^2},\qquad 
\overline{m}^2\equiv R^2 m^2.
\end{equation}
The critical crossover function $F_\chi(\widetilde{t}\,)$ 
can be obtained by rewriting Eq.~(\ref{sapo}) 
for $h=0$ in terms of 
$\widetilde{t}$ and $\widetilde{\chi}$, and taking the limit 
$R\to\infty$ (thus $\overline{G}\to 0$) with $\overline{u}$ fixed:
\begin{equation}
\widetilde{t} = N\widetilde{\chi}^{-1} + {N\overline{u}\over 6}
\lim_{R\to \infty}\int {d^dk\over (2\pi)^d} 
{N\widetilde{\chi}^{-1} \over \overline{\Pi}
( \overline{\Pi} + \overline{G} N\widetilde{\chi}^{-1}  ) },
\label{chisceq}
\end{equation}
where $\overline{\Pi}\equiv R^2 \Pi$.
An analysis of the integral in Eq.~(\ref{chisceq}) shows that,
for $2 < d < 4$,  the limit
exists. 
It depends only on the following property
of $\overline{\Pi}$:
\begin{equation}
\lim_{q\to \infty} 
\;\left[ \lim_{R\to \infty}\overline{\Pi}(q/R,R)\right]
= b_{\infty}, \label{proppi1} 
\end{equation}
where $b_\infty$ is a non-vanishing constant.
Its explicit value is not relevant, 
since a change of this constant can 
be reabsorbed in a change of normalization for $\widetilde t$.
Therefore our results are universal, apart from a 
rescaling of $\widetilde t$ and $\widetilde\chi$,
for a large class of Hamiltonians satisfying the above condition.
{}From Eq. (\ref{chisceq}) one finally obtains
\begin{eqnarray}
&&\widetilde{t} \,{F_\chi(\widetilde{t}\,)\over N} 
= K  +  L_d 
\left( {F_\chi(\widetilde{t}\,)\over N}\right)^{2-d/2} ,
\label{finchieq}\\
&&K=1 + {N\overline{u}\over 6b_\infty^2},\qquad
L_d = - {\Gamma(1 - d/2)\over 6 (4\pi)^{d/2}}.
\nonumber 
\end{eqnarray}
Eq.~(\ref{finchieq})
is well defined in the large-$N$ limit after proper rescalings
in $N$ of the fields (i.e. $\phi\to \sqrt{N}\phi$) and couplings
(i.e. $\overline{u}\to \overline{u}/N$).
The expected asymptotic behaviors are reproduced:
\begin{equation}
{F_\chi(\widetilde{t}) \over N} \sim 
\cases{
    \widetilde{t}^{-2/(d-2)} 
    (1 + c_1 \widetilde{t}^{\Delta_{\rm sr}}+...) 
                    & for $\widetilde t \to 0$,\cr
    \widetilde{t}^{-1} 
    (1 + c_2 \widetilde{t}^{-\Delta_{\rm g}}+...) 
                    & for $\widetilde t\to\infty$,
      }
\end{equation}
where $\Delta_{\rm sr} = (4-d)/(d-2)$ and 
$\Delta_{\rm g} = (4-d)/2$ are the leading 
correction-to-scaling exponents related to the nonclassical
short-ranged and 
the Gaussian fixed point respectively.
The coefficient of the $\widetilde{t}^{\Delta_{\rm sr}}$ correction 
in the expansion around $\widetilde{t}=0$ 
(and in general also the other coefficients of the expansion)
can be obtained
by performing the appropriate limit of the nonuniversal
Wegner expansion.
We stress that the dependence on the
bare coupling $\overline{u}$ in Eq.~(\ref{finchieq}) can be eliminated
by a rescaling and a redefinition of $\widetilde t$.
Moreover, the same equation (modulo the above mentioned
rescalings) can be obtained starting from the $N$-vector 
(nonlinear $\sigma$) model.

The analysis of experimental data in the crossover region is
usually performed by
introducing effective critical exponents.
One can define $\gamma_{\rm eff}$
by the logarithmic derivative of $F_\chi(\widetilde{t})$. In 
three dimensions 
\begin{equation}
\gamma_{\rm eff} \equiv - {d \ln F_\chi\over d \ln \widetilde{t}}=
1 + \left(1 + c_\gamma \widetilde{t} \,\right)^{-1/2},
\label{gaeff}
\end{equation}
where $c_\gamma= 4K/L_3^2$.
$\gamma_{\rm eff}(\widetilde{t}\,)$ is universal
apart from a trivial rescaling of $\widetilde{t}$.
Analogously one may define $\nu_{\rm eff}$ and find
\begin{equation}
\nu_{\rm eff} \equiv - {1\over 2} {d \ln F_{\xi^2}\over d \ln \widetilde{t}}=
{\gamma_{\rm eff}\over 2}.
\label{xieff}
\end{equation}

{}From Eqs.~(\ref{sapo}) and  (\ref{M-h-beta-relation})
one can derive an
equation of state relating the rescaled variables $\widetilde t$,
$\widetilde{h}$,  and 
$\widetilde M$ in the critical crossover limit. 
Simple calculations,
involving the same integral of Eq.~(\ref{chisceq}),
lead to the equation
\begin{equation}
{\widetilde{M}^2\over 6 N} + 
\widetilde{t}  =
K{\widetilde{h}\over \widetilde{M}} +
L_d \left( {\widetilde{h}\over \widetilde{M}}\right)^{d/2-1}
\label{creqstate} 
\end{equation}
which turns out  to be  universal apart from trivial rescalings
of $\widetilde{t}$, $\widetilde{h}$, and $\widetilde{M}$.
Moreover it reproduces the correct asymptotic behaviours:
\begin{equation}
\widetilde{h} \sim \widetilde{M}^{\delta} 
\left( 1 + c \widetilde{t}\,\widetilde{M}^{-1/\beta}
  \right)^\rho
  \left[ 1 + O\left(\widetilde{M}^{\Delta/\beta}\right)\right],
\label{asbeheqst-shortrange}
\end{equation}
where  in the nonclassical limit ($\widetilde{t}\to 0$ and 
$\widetilde{h}\to 0$)
$\delta = (d+2)/(d-2)$, $\beta=1/2$, and $\rho=2/(d-2)$.
In the classical limit
$\delta=3$, $\beta=1/2$ and $\rho=1$.
Setting $\widetilde{t}=0$ in Eq.~(\ref{creqstate}),
one can derive the crossover function $F_M(\widetilde{h})$ defined 
in Eq.~(\ref{scalfunc}).
In three dimensions the corresponding
effective exponent $\delta_{\rm eff}$ is given by
\begin{equation}
\delta_{\rm eff}\equiv 
{d\ln \widetilde{h}\over d\ln F_M}=
3 + 2 \left(1 + {c_\gamma\over6}\, { F_M^2\over N} \right)^{-1/2}.
\label{deeff}
\end{equation}

We can also consider the short-range version of the model and show that
the function 
$F_\chi^{\rm sr}(\widetilde{t}\,)\equiv\chi (Nu)^{2/(4-d)}$
satisfies Eq.~(\ref{finchieq}) with $b_\infty\to\infty$. 
The same arguments apply to all  other critical crossover functions.
This confirms that the critical crossover functions are universal,
i.e. independent of the mechanism driving $G$.

In the symmetric phase, we can define the
zero-momentum four-point coupling $g$ as 
\begin{equation}
g = -{3N\over N+2} {\chi_4\over \chi^2\xi^d},
\end{equation}
where $\chi_4$ is the connected four-point correlation
function at zero momentum.
In the large-$N$ limit~\cite{ONgr}
\begin{equation}
N g = \widetilde{m}^{d-4}
\left[ 1 + {N\overline{u}\over 6}
\int {d^dk\over (2\pi)^d} 
{ 1 \over 
( \overline{\Pi} + \overline{G}\widetilde{m}^2 )^2 }\right]^{-1},
\label{fcr}
\end{equation}
where 
$\widetilde{m}^2\equiv \widetilde{\xi}^{\,-2}$.
In the critical crossover limit the integral depends only
on $b_\infty$. We obtain
\begin{equation}
{g(\widetilde{m})\over g(0)} = {1\over 1 + c_g \widetilde{m}^{4-d}},
\label{fcr2}
\end{equation}
where $Ng(0)=6 (4\pi)^{d/2}/\Gamma(2-d/2)=Ng^*$ 
is the nonclassical critical value of $g$,
and $c_g = 2K/(d-2)L_d$.
The effective critical exponent associated with $g(\widetilde{m})$
is 
\begin{equation}
\psi_{\rm eff}(g)\equiv {d \ln g(\widetilde{m})\over d\ln\widetilde{m}}
= (d-4)\left[ 1 - {g\over g^*}\right].
\label{feeff}
\end{equation}
In the same limit
Eqs.~(\ref{gaeff}), (\ref{xieff}) and (\ref{fcr2}) imply also 
\begin{equation}
\gamma_{\rm eff} (g) = 2\nu_{\rm eff}(g) = 1 + {4-d\over d-2} {g\over g^*}.
\label{gafff}
\end{equation}
Eqs.~(\ref{feeff}) and (\ref{gafff}) are now 
independent of $\overline{u}$ and $b_\infty$.
Notice that only in the nonclassical critical
limit ($\widetilde{m}\to 0$), $\psi_{\rm eff}(g)\to 0$
and therefore the corresponding hyperscaling relation is
satisfied.
This fact is not unexpected because hyperscaling 
is not satisfied at the Gaussian fixed
point,
where $g\sim (T-T_c)^{(4-d)/2}$ for $T\to T_c$. 

\begin{figure}
\vspace*{-1.2cm} \hspace*{0cm}
\begin{center}
\epsfxsize = 0.42\textwidth
\vspace{-2.0cm}
\leavevmode\epsffile{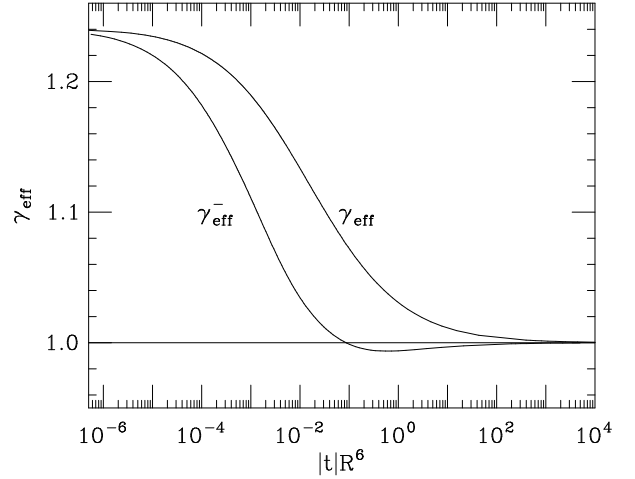}
\end{center}
\vspace*{0.0cm}
\caption{Effective susceptibility exponent as a function of 
$\widetilde{t}$ for the high- ($\gamma_{\rm eff} $) and 
low- ($\gamma_{\rm eff}^- $) temperature phase of the three-dimensional 
Ising model. Here $t$ is the reduced temperature
for the model with Hamiltonian (\protect\ref{lhamB}) and $R$ is 
defined in Eq. (\protect\ref{defR}).
}
\label{fig}
\end{figure}

One can now easily verify
that $\psi_{\rm eff}(g) =\beta(g)/g$ where $\beta(g)$ is the Callan-Symanzik 
$\beta$-function in the continuum $\phi^4$ theory. Analogously 
the high-temperature exponents
$\gamma_{\rm eff}(g)$ and $\nu_{\rm eff}(g)$ are related to the  standard 
RG functions $\gamma(g)$ and $\nu(g)$ (see e.g. Ref.~\cite{B-B})
through the relations \cite{alphaeff}
\begin{eqnarray}
&&{\gamma_{\rm eff}(g)\over\nu_{\rm eff}(g)}=
{\gamma(g)\over\nu(g)}, \\
&&\nu(g)
  \beta(g) {d\gamma_{\rm eff}\over dg} = \gamma(g)- \gamma_{\rm eff}(g).
\end{eqnarray}
In the large-$N$ limit 
\begin{equation}
2 \gamma^{-1}(g) = \nu^{-1}(g) = 2 + (d-4) g/g^*.
\end{equation}

Using the field theoretical approach \cite{B-B-l,B-B,B-B-M-N}, 
one can compute the effective exponents in three-dimensional 
$O(N)$ models. Results for the high-temperature phase are reported in 
Refs. \cite{B-B-l,B-B}. 
We extended the computation \cite{LT} to the low-temperature 
phase computing $\gamma_{\rm eff}^-$ for $N=1$.
The resulting curves are plotted in Fig.~\ref{fig}.
We stress that, apart from the small error of the resummation procedure 
--- it should be well below 1\% ---
the curves in Fig.~\ref{fig} represent the universal critical crossover
exponents. Thus experimental and numerical data in the crossover
region should approach these curves in the appropriate limit
(modulo a rescaling of $\widetilde{t}\;$).
In this perspective it is possible to understand the lack of universality
of the results of Ref.~\cite{A-P-K-S}: universality is recovered
only in the limit $u\to 0$. For finite values of $u$ 
one expects corrections to scaling that eventually disappear as $u\to 0$.
Indeed the comparison with
the experimental data for fluids and liquid mixtures \cite{A-P-K-S} improves  
as the effective parameter $u$ decreases.

Measurements in the critical crossover limit are difficult and accurate
results are scarce. Experiments on micellar solutions~\cite{C-D} 
observed exponents that were very far from the expected 
Ising values. The exponent $\gamma$ was even lower
than  the classical value $\gamma=1$.
Fisher~\cite{Fisher-prl} interpreted the data as a crossover effect,
suggesting a standard scaling description.
In order to explain the data, this interpretation would require 
$\gamma_{\rm eff}$ to be non-monotonic in the symmetric phase.
But, as Fig.~\ref{fig} shows, this is not the case for 
the critical crossover function $\gamma_{\rm eff}$.
Therefore,
as already observed by Bagnuls and Bervillier~\cite{B-B-prlf},
the results of \cite{C-D} cannot be explained 
in terms of universal crossover functions.

On the other hand, it is interesting to note that 
$\gamma_{\rm eff}^-(\widetilde{t}\,)$ is non-monotonic.
In three dimensions the effect is rather small (see Fig. \ref{fig}).
The minimum value of $\gamma_{\rm eff}^-(\widetilde{t}\,)$
is $\gamma_{\rm eff}^-(\widetilde{t}\,)|_{\rm min}\approx 0.994$, so that
a non-monotonic behavior can hardly be seen.
This type of behavior had already been observed numerically in 
two dimensions \cite{L-B-B-prl}: in this case, however, the effect 
was much larger.
The non-monotonicity of $\gamma_{\rm eff}^-$ can be predicted analytically
by calculating the first correction to the mean-field behavior
in the low-temperature phase. 
One can indeed show that $\gamma_{\rm eff}^-(\widetilde{t}\,)$
is increasing for $|\widetilde{t}|\rightarrow \infty$.
For instance,
let us consider the long-range Ising model introduced in Ref.~\cite{M-B}
and defined by the Hamiltonian
\begin{equation}
{\cal H}=-\sum_{i,j} J(\vec{x}_i-\vec{x}_j) s_i s_j,
\label{lhamB}
\end{equation}
where $J(\vec{x}) = c R_m^{-d}$ for 
$|\vec{x}|\leq R_m$ and $J(\vec{x}) = 0 $ otherwise.
Setting  $t = (\beta_c - \beta)/\beta_c$ and 
$\widetilde{t} = t R^{2d/(4-d)}$, one finds for $2<d<4$
\begin{equation}
\gamma_{\rm eff} = 1 + C |\widetilde{t}|^{(d-4)/2} + O(|\widetilde{t}|^{d-4}),
\label{gammaeff}
\end{equation}
where the constants $C$ are given by: 
\begin{eqnarray}
C^+ &=& {\Gamma(3-d/2) \over 2^{d-1} \pi^{d/2} (d-2)}, \\
C^- &=& - 2^{2-d/2}\left[1 - 3\cdot2^{d-5}(d-2)\right] \, C^+.
\end{eqnarray}
in the high- and low-temperature phases respectively. 
Notice that for $d = 3$, 
$C^-$ is negative so that $\gamma_{\rm eff}^-(\widetilde{t}\,)$ 
cannot be monotonic.  In two dimensions
the subleading corrections behave as $\widetilde{t}^{-1}\log \widetilde{t}$, 
again with a negative coefficient. In three dimensions, in Eq. (\ref{gammaeff}),
logarithms appear at next next-to-leading order. 

We have also performed a high-statistics 
Monte Carlo simulation of 
a $d=3$ model of self-avoiding walks with long-range 
interactions~\cite{C-C-P-R-V}. We obtain 
evidence of the existence of a universal critical crossover scaling 
in the large-$R$ limit. The resulting universal curves turn out to be
in agreement with the two-parameter model predictions~\cite{M-N,B-N}, 
confirming the general arguments.



\begin{references}

\bibitem{C-D}
M.~Corti, V.~Degiorgio,
Phys.\ Rev.\ Lett.\ {\bf 55}, 2005 (1985).

\bibitem{Fisher-prl} 
M.~E.~Fisher, Phys.\ Rev.\ Lett.\ {\bf 57}, 1911 (1986).

\bibitem{A-P-K-S} 
M.~A.~Anisimov, A.~A.~Povodyrev,
V.~D.~Kulikov, and J.~V.~Sengers,
Phys.\ Rev.\ Lett.\ {\bf 75}, 3146 (1995).

\bibitem{F-L} M.~E.~Fisher,
B.~P.~Lee, Phys.\ Rev.\ Lett.\ {\bf 77}, 3561 (1996).


\bibitem{Ginzburg} V.~L.~Ginzburg,
Fiz.\ Tverd.\ Tela {\bf 2}, 2031 (1960).

\bibitem{D-B} H.-P.~Deutsch, K.~Binder,
J.\ Phys.\ (France) II {\bf 3}, 1049 (1993).

\bibitem{S-F} P.~Seglar and M.~E.~Fisher,
J.\ Phys.\ {\bf C 13}, 6613 (1980). 

\bibitem{N-B} J.~F.~Nicoll and  J.~K.~Bhattacharjee,
Phys.\ Rev.\ {\bf B 23}, 389 (1981).

\bibitem{B-B-l} C.~Bagnuls and  C.~Bervillier,
J. Phys. Lett. (Paris) {\bf 45},
L-95 (1984).

\bibitem{N-A} J.~F.~Nicoll, P.~C.~Albright,
Phys.\ Rev.\ {\bf B 31}, 4576 (1985).

\bibitem{B-B} C.~Bagnuls,  C.~Bervillier,
Phys.\ Rev.\ {\bf B 32}, 7209 (1985).

\bibitem{B-B-M-N} 
C.~Bagnuls, C.~Bervillier, D. I. Meiron, and B. G. Nickel,
Phys.\ Rev.\ {\bf B 35}, 3585 (1987).

\bibitem{C-A-S} Z.~Y.~Chen, P.~C.~Albright,
and J.~V.~Sengers, Phys.\ Rev.\ {\bf A 41}, 3161 (1990).

\bibitem{A-K-S-T} 
M.~A.~Anisimov, S.~B.~Kiselev, J.~V.~Sengers and S.~Tang,
Physica {\bf A 188}, 487 (1992).

\bibitem{B-K} 
M.~Y.~Belyakov, S.~B.~Kiselev, 
Physica {\bf A 190}, 75 (1992).

\bibitem{notawe} Even 
the sign of the leading correction depends on the physical system,
see e.g. A.~J.~Liu and M.~E.~Fisher,
J.\ Stat.\ Phys.\ {\bf 58}, 431 (1990).

\bibitem{B-B-prl} C.~Bagnuls, C.~Bervillier,
Phys.\ Rev.\ Lett.\ {\bf 76}, 4094 (1996).

\bibitem{A-P-K-S-2} 
M.~A.~Anisimov, A.~A.~Povodyrev,
V.~D.~Kulikov and  J.~V.~Sengers,
Phys.\ Rev.\ Lett.\ {\bf 76}, 4095 (1996).

\bibitem{L-B-B-pre} 
E.~Luijten, H.~W.~J.~Bl\"ote and K.~Binder,
Phys.\ Rev.\ {\bf E 54}, 4626 (1996).

\bibitem{L-B-B-prl} 
E.~Luijten, H.~W.~J.~Bl\"ote and K.~Binder,
Phys.\ Rev.\ Lett.\ {\bf 79}, 561 (1997);
Phys.\ Rev.\ {\bf E 56}, 6540 (1997).

\bibitem{M-B} K.~K.~Mon and K.~Binder,
Phys.\ Rev.\ {\bf E 48}, 2498 (1993).

\bibitem{M-N} M.~Muthukumar and B.~G.~Nickel,
J. Chem. Phys. {\bf 80}, 5839 (1984).

\bibitem{Sokal} 
A.~D.~Sokal, Europhys.\ Lett.\ {\bf 27}, 661 (1994).

\bibitem{B-N} P.~Belohorec, B.~G.~Nickel, 
Guelph Univ. report (1997).

\bibitem{ONgr}
M.~Campostrini, A.~Pelissetto,
P.~Rossi, E.~Vicari, Nucl.\ Phys.\ {\bf B459}, 207 (1996).


\bibitem{alphaeff}
Notice that the exponent $\alpha_{\rm eff}$, defined  as the logarithmic
derivative of the specific heat, does not have a universal crossover
limit (and thus is not directly related to RG functions) due
to the presence of the analytic background.

\bibitem{LT} 
In Ref. \cite{B-B-M-N} $\gamma^{-}_{\rm eff}$ was already computed
in a neighbourhood of the critical point. However some of the perturbative
series used in the calculation were incorrect. 
Our computation uses the correct 
values that have been derived from the results  of 
Refs.~\cite{H-D,G-Z}. In particular, in Table 3 of Ref. \cite{B-B-M-N},
one should correct the five-loop coefficients of $X(g)$ and $S(g)$ as 
follows:
$X_5 =  0.020211485$; $S_5 = 0.0433684818$. 


\bibitem{H-D} 
F.~J.~Halfkann and V.~Dohm,
Z. Phys. {\bf B 89}, 79 (1992).

\bibitem{G-Z} R.~Guida, J.~Zinn-Justin,
Nucl.\ Phys.\ {\bf B 489}, 626 (1997).


\bibitem{B-B-prlf} C.~Bagnuls, C.~Bervillier,
Phys.\ Rev.\ Lett.\ {\bf 58}, 435 (1987).
 
\bibitem{C-C-P-R-V} S.~Caracciolo, M.~S.~Causo,
A.~Pelissetto, P.~Rossi, and E.~Vicari, in preparation.




\end{references}
\end{document}